\documentclass[prd,twocolumn,showpacs,aps,epsbox,groupedaddress,eqsecnum]{revtex4}
\usepackage[dvips]{graphicx}
\usepackage{bm}
\usepackage{color}
\usepackage{epsfig}
\usepackage{here}
\usepackage{subfigure}
\usepackage{amsmath}

\def\beq{\begin{equation}}
\def\eeq{\end{equation}}
\def\bea{\begin{eqnarray}}
\def\eea{\end{eqnarray}}

\begin{document}
\draft
\title{Can we identify massless braneworld black holes by observations?}

\author{Masashi Kuniyasu}
\email{u003wa@yamaguchi-u.ac.jp}
\author{Keitaro Nanri}
\author{Nobuyuki Sakai}
\email{nsakai@yamaguchi-u.ac.jp}
\affiliation{Faculty of Science, Yamaguchi University, Yamaguchi 753-8512, Japan}
\author{Takayuki Ohgami}
\email{t-ohgami@daido-it.ac.jp}
\affiliation{Department of Physics, Daido University, Nagoya 457-8530, Japan}
\author{Ryosuke Fukushige}
\affiliation{Graduate School of Science and Engineering, Kagoshima University, Kagoshima 890-0065, Japan}
\author{Subaru Komura}
\affiliation{Graduate School of Engineering, Nagoya University, Nagoya 464-8603, Japan}

\begin{abstract}
For an extension of the previous work on gravitational lensing by
massless braneworld black holes, we investigate their microlensing phenomena and
shadows and discuss how to distinguish them from standard Schwarzschild black holes and Ellis wormholes.
Microlensing is known as the phenomenon in which luminosity amplification appears when a bright object passes behind a black hole or another massive object.
We find that, for the braneworld black hole as well as for the Ellis wormhole, there appears luminosity reduction just before and after the amplification.
This means that observation of such a reduction would indicate the lens object is either a braneworld black hole or a wormhole, though it is difficult to distinguish one from the other by microlensing solely.
Therefore, we next analyze the optical images, or shadows of the braneworld black hole surrounded by optically thin dust, and compare them to those of the Ellis wormhole.
Because the spacetime around the braneworld black hole possesses unstable circular orbits of photons, a bright ring appears in the image, just as in Schwarzschild spacetime or in the wormhole spacetime.
This indicates that the appearance of a bright ring does not solely confirm a braneworld black hole, a Schwarzschild, nor an Ellis wormhole.
However, we find that only for the wormhole is the intensity inside the ring larger than that the outsider intensity.
Therefore, with future high-resolution observations of microlensing and shadows together, we could identify the braneworld black holes if they exist.
\end{abstract}

\maketitle

\section{Introduction}
LIGO has recently detected gravitational waves, which are supposed to be signals of coalescence of two black holes \cite{LIGO}.
It is also expected that direct observation of black holes will be achieved by very long baseline interferometry  observations in the near future \cite{VLBI}.
Those advancements in observational technology to detect black holes will also give us a chance to discover exotic compact objects such as boson stars \cite{BS1}, gravastars \cite{GVS1}, wormholes \cite{WH1}, non-Abelian black holes \cite{NABH}, and braneworld black holes \cite{EHM,BMD,Kanti,BB}.

To detect such objects, it is necessary to understand theoretical predictions for observation in advance.
For this purpose, observational consequences of boson stars \cite{BS2}, gravastars \cite{GVS2,GVS3}, wormholes \cite{WH2,WH3,WH4,WH5,WH6,OS}, and braneworld black holes \cite{lens,MM,ES} have been studied recent years.

Among many models of braneworld black holes, massless black holes \cite{EHM}, in which the curvature is produced only by a tidal effect \cite{BMD}, are observationally important because their gravitational lensing effects are characteristic and discriminative \cite{MM,ES}.
In this paper, we study gravitational lensing by massless braneworld black holes in more detail.
Specifically, we study their microlensing and shadows, and discuss whether we can distinguish them from standard Schwarzschild black holes and Ellis wormholes by radio or electromagnetic observations.

This paper is organized as follows.
In Sec. II, we introduce three spacetime models to be studied: a massless braneworld black hole, a Schwarzschild black hole, and an Ellis wormhole.
In Sec. III, we derive null geodesic equations and analyze photon trajectories;
we also discuss deflection angles for the three models.
In Sec. IV, we investigate observational consequences of microlensing phenomena. Concretely,
we solve null geodesic equations numerically to obtain the images of an optical source object behind a lens object for the three models;
we also calculate the light curves, that is, the time variation of the total luminosity. 
In Sec. V, we derive stationary solutions of dust fluid for the three models by solving the energy-momentum conservation, and we investigate their images (i.e., shadows) by solving the radiative transfer equation as well as the null geodesic equations.
Finally, in Sec. VI, we discuss whether we can identify massless braneworld black holes by electromagnetic observations.

We use the units of $c=1$ in this paper.

\section{Spacetime models}

We start with the general expression of the metric for spherically symmetric and static spacetimes
\beq\label{metric}
ds^2=-A(r)dt^2+B(r)dr^2+C(r)(d\theta^2+\sin^2\theta d\psi^2).
\eeq
Bronnikov {\it et al.} \cite{BMD} studied a general class of braneworld black holes in five-dimensional Einsten equations;
the metric functions for massless black holes are generally expressed as 
\bea
&&A=1-\frac{{r_g}^2}{r^2},~~~
B^{-1}=A\left(1+{D-r_g\over\sqrt{2r^2-r_g^2}}\right),\nonumber\\
&&C=r^2,
\eea
where $r_g$ and $D$ are constants and $r_g$ corresponds to an event horizon.

In this paper, for a simplest model in this class, we suppose $D=r_g$, that is,
\beq\label{5S}
A=B^{-1}=1-\frac{{r_g}^2}{r^2},~~~
C=r^2.
\eeq
This solution is identified as the five-dimensional Schwarzschild metric or the Reissner-Nordstr\"om metric with pure imaginary charge.

For comparison, we also analyze two models. One is the standard Schwarzschild spacetime
\beq\label{4S}
A=B^{-1}=1-\frac{r_g}{r},~~~
C=r^2,
\eeq
where $r_g$ is also a constant, which corresponds to the event horizon.
Because $(r_g/r)^2$ decreases faster than $r_g/r$ as $r$ increases, the strength of gravity in the five-dimensional Schwarzschild spacetime is smaller than the standard Schwarzschild spacetime under fixed $r_g$.

The other reference spacetime is the Ellis wormhole
\begin{equation}\label{EWH}
A=B=1,~~~C=r^2+a^2,
\end{equation}
where $a$ is the throat radius of the wormhole.
Note that $r=0$ corresponds to the throat and $r<0$ corresponds to ``the other side."

\section{Photon trajectories}
\subsection{Null geodesic equations and effective potential}

In this section, we derive null geodesic equations for the three models. We denote the affine parameter and the null vector by $\lambda$ and $k^\mu\equiv dx^\mu/d\lambda$, respectively; then, the geodesic equations are generally given by
\begin{equation}\label{geodesic}
\frac{dk^{\mu}}{d\lambda}+{\Gamma}^{\mu}_{\nu \rho}k^{\nu}k^{\rho}=0,
~~~{\rm with}~~~
k_{\mu}k^{\mu}=0.
\eeq
For the general metric form (\ref{metric}), we write down the geodesic equations (\ref{geodesic}) in the $\theta =\pi/2$ as
\begin{equation}\label{GEtp}
\frac{d}{d\lambda}\left(Ak^{t}\right)=0,\,\,\,\frac{d}{d\lambda}\left(C k^{\varphi}\right)=0.
\end{equation}
\beq\label{GEr}
\frac{d}{d\lambda}\left({k^{r}\over A}\right)+{A'\over2}(k^t)^2+{A'\over2A^2}(k^r)^2-{C'\over2}(k^\varphi)^2=0,
\eeq
\begin{equation}\label{null}
-A\left(k^{t}\right)^2+B\left( k^{r} \right)^2 +C\left(k^\varphi\right)^2=0,
\end{equation}
where $'\equiv d/dr$. Because (\ref{GEr}) is also derived by (\ref{GEtp}) and (\ref{null}), we do not have to solve it.
Equations (\ref{GEtp}) are integrated as
\begin{equation}\label{EL}
Ak^t={\rm const}\equiv E ,\,\,\,\,Ck^\varphi={\rm const}\equiv L,
\end{equation}
and then (\ref{null}) becomes
\beq\label{null2}
\left({dr\over d\lambda}\right)^2+V(r)=E^2,~~~
V(r)\equiv\frac{A}{C}L^2.
\eeq
Here, the effective potential $V(r)$ is introduced to discuss light trajectories by analogy with Newtonian mechanics.

\begin{figure}
\centering
\subfigure[Braneworld black hole. ]{
\includegraphics[scale=0.35]{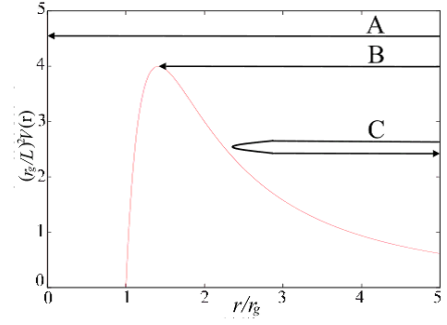}
\label{aa}}
\subfigure[Ellis wormhole.]{
\includegraphics[scale=0.35]{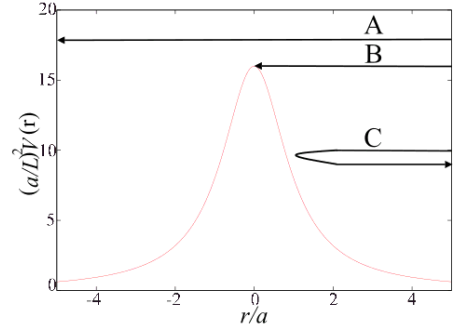}
\label{aaa}}
\caption{Effective potentials of null geodesics equations around (a) the braneworld black hole and (b) the Ellis wormhole.
The maximum points correspond to unstable circular orbits of photons.}
\label{aaaa}
\end{figure}

Figure \ref{aa} shows the effective potential for the braneworld black hole,
\beq\label{VrBBH}
V(r)={L^2\over r^2}\left(1-{r_g^2\over r^2}\right).
\eeq
As in the case of the Schwarzschild black hole, there are unstable circular orbits of photons, which play an important role in microlensing and generating shadows.
The trajectories are classified into three types: A, B, and C. Type C represents the photon that is refracted and goes to infinity.
Type A represents the photon that goes to singularity. Type B represents the photon that approaches and winds many times in the vicinity of the unstable circular orbit.

Figure \ref{aaa} shows the effective potential for the Ellis wormhole,
\beq\label{VrEWH}
V(r)={L^2\over r^2+a^2}.
\eeq
The region $r < 0$ corresponds to the opposite side of the wormhole.
The maximum point at $r=0$ corresponds to unstable circular orbits of photons.

For reference, the effective potential for the Schwarzschild spacetime is given by
\beq\label{VrSBH}
V(r)={L^2\over r^2}\left(1-{r_g\over r}\right).
\eeq
\subsection{Photon trajectories}
Using the effective potential derived in the previous subsection, we can discuss qualitative properties of photon trajectories in the three types of spacetimes. 
For the braneworld black hole, the trajectories are classified into three types: A, B, and C. 
Type A represents the photon that passes across the unstable circular orbit and goes into singularity.
Type C represents the photon that passes outside the unstable circular orbit and goes to infinity.
Type B represents the photon that approaches and winds many times in the vicinity of the unstable circular orbit.

To investigate those trajectories more closely, we numerically solve the null geodesic equations.
We define the rectangle coordinates as 
\beq\label{xyz}
x=\sqrt{C}\cos\varphi\sin\theta,~y=\sqrt{C}\sin\varphi\sin\theta,~z=\sqrt{C}\cos\theta.
\eeq
Figure 2(a) shows some trajectories of photons on the $z=0$ plane that reach the observer at $x=300r_g,~y=0$.
Labels A, B, and C correspond to those in the effective potential in Fig. 1.
We confirm that there are really three types of trajectories, as we discussed above with the effective potential.

For reference, we also plot photon trajectories around Schwarzschild black hole and the Ellis wormhole in Figs. 2(b) and in 2(c), respectively.
For each case, the trajectories are classified into three types, depending on whether the photon passes across the unstable circular orbit or not. In the case of the Ellis wormhole, type A represents the photon that comes from the other side of the wormhole.

\begin{figure}[H]
\centering
\subfigure[Braneworld black hole]{
\includegraphics[scale=0.24]{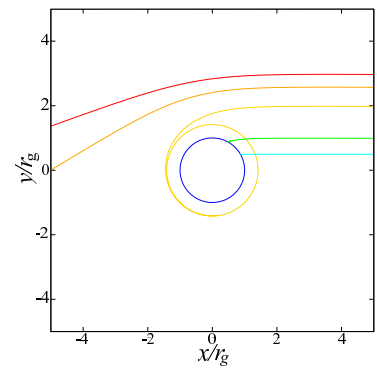}
\label{schst}}
\subfigure[Schwarzschild black hole]{
\includegraphics[scale=0.24]{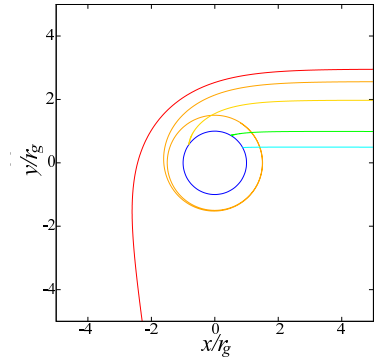}
\label{brast}}
\subfigure[Ellis wormhole]{
\includegraphics[scale=0.29]{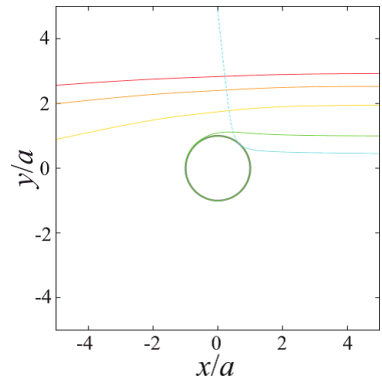}
\label{ewst}}
\caption{Photon trajectories around (a) the braneworld black hole, (b) the Schwarzschild black hole, and (c) the Ellis wormhole.
The coordinates $(x,y)$ are defined by (\ref{xyz}) on the $z=0~(\theta=\pi/2)$ plane.
We suppose an observer at $x=300r_g$ (or $x=300a$), $y=0$, at which all the trajectories end up.
The blue circle denotes the surface of lens objects. 
In (a) and (b) types A, B, and C correspond to those in the effective potential in Fig.\ 1.
The dashed line in (c) represents the trajectory on the other side of the wormhole $(r<0)$.
}
\label{st}
\end{figure}

\subsection{Refraction angles}

Before performing numerical analysis, it is instructive to review the analytic formula of refraction angles for the three models.
As shown in Fig.\ \ref{fig:gl}, we approximate the geodesic by its two tangent lines at the observer and at the source in a fictitious Euclidean plane.
The refraction angle $\delta$ is defined as the angle between the two tangent lines.
The impact parameter $\alpha$ is defined as the distance between the lens object's center and the intersection of the two tangent lines.
If the observer is sufficiently far from the object, the impact parameter is approximated by
\beq\label{alphaLE}
\alpha=\frac LE,
\eeq
where $E$ and $L$ are integral constants defined by (\ref{EL}), which can be interpreted as the energy and the angular momentum, respectively.

\begin{figure}
\centering
\includegraphics[scale=0.4]{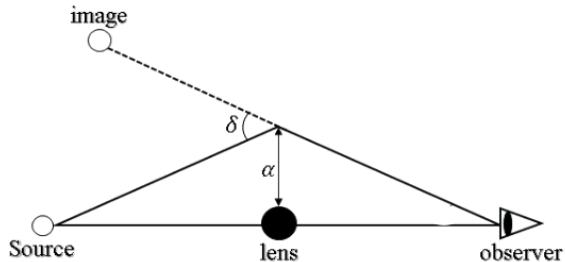}
\caption{Definition of $\delta$ and $\alpha$ in lensing phenomena.
We approximate the geodesic by its two tangent lines at the observer and at the source in a fictitious Euclidean plane.
The refraction angle $\delta$ is defined as the angle between the two tangent lines.
$\alpha$ denotes the impact parameter.
}
\label{fig:gl}
\end{figure}

In the weak field limit, the analytic formulas of $\delta$ have been obtained in the literature. The leading term for each model is as follows.:
\begin{itemize}
\item Schwarzschild black hole:
\begin{equation}
\delta =\frac{2r_g}{\alpha}.
\end{equation}
\item Braneworld black hole \cite{MM}:
\begin{equation}
\delta =\frac{3\pi{r_g}^2}{4\alpha^2}.
\end{equation}
\item Ellis wormhole \cite{WH3,WH4}: 
\begin{equation}\label{deltaEH}
\delta =\frac{\pi a^2}{4\alpha^2}.
\end{equation}
\end{itemize}
The refraction angle for the Ellis wormhole was originally derived by Dey and Sen \cite{WH3}.
However, Nakajima and Asada \cite{WH4} showed that the previous result breaks down at the next-to-leading order though the leading term (\ref{deltaEH}) itself is valid. These arguments were also reviewed in Ref.\cite{WH5}.

It should be noted that both deflection angles of the braneworld black hole and of the Ellis wormhole are proportional to $\alpha^{-2}$, while that of the Schwarzschild black hole is proportional to $\alpha^{-1}$.
This indicates that the braneworld black hole and the Ellis wormhole may exhibit similar features in microlensing phenomena, which will be checked in the next section.

\section{Gravitational microlensing}

\begin{figure}[b]
\centering
\includegraphics[scale=0.33]{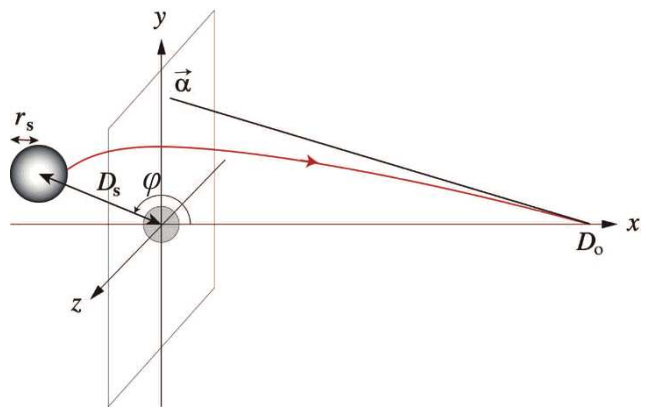}
\caption{Setting of our numerical analysis of gravitational lens effects in the rectangular coordinate system.
We place the lens object's center at the origin and the observer at $x=D_o$ on the $x$ axis.
We suppose that a source object is moving behind the lens object on the $z=0$ plane.
We denote the distance between the source object's center and the lens object's center by $D_s$ and the radius of the source by $r_s$.
The position of the source is characterized by $\varphi$, which is defined as the angle between the direction of the source object's center and the the direction of the observer.
The image $\vec\alpha=(\alpha_y,~\alpha_z)$ is defined as the intersection of the $x=0$ plane with the tangent to the ray at the observer.}
\label{fig:zg}
\end{figure}
Figure 4 shows the setting of our numerical analysis of gravitational lens effects in the rectangular coordinate system defined by (\ref{xyz}).
We place the lens object's center at the origin and the observer at $x=D_o$ on the $x$ axis.
We suppose that a source object is moving behind the lens object on the $z=0$ plane.
We denote the distance between the source object's center and the lens object's center by $D_s$ and the radius of the source by $r_s$.
The position of the source is characterized by $\varphi$, which is defined as the angle between the direction of the source object's center and the the direction of the observer.
The image $\vec\alpha=(\alpha_y,~\alpha_z)$ is defined as the intersection of the $x=0$ plane with the tangent to the ray at the observer.
In our numerical calculation below we set $D_o=300r_g,~D_s=10r_g$ or $10a$, and $r_s=r_g$.

Figure 5 shows an example of the images of the source object $\vec\alpha=(\alpha_y,~\alpha_z)$ around the braneworld black hole.
We display four snapshots when $\varphi$=$130^\circ,~150^\circ,~170^\circ$, and $180^\circ$.
The image becomes distorted as $\varphi$ increases.
The light-colored images correspond to geodesics that wind in the vicinity of unstable circular orbits.
Einstein rings appear when $\varphi$=$180^\circ$. 

\begin{figure}[H]
\begin{center}
\subfigure[$\varphi$=$130^\circ$]{
\includegraphics[scale=0.23]{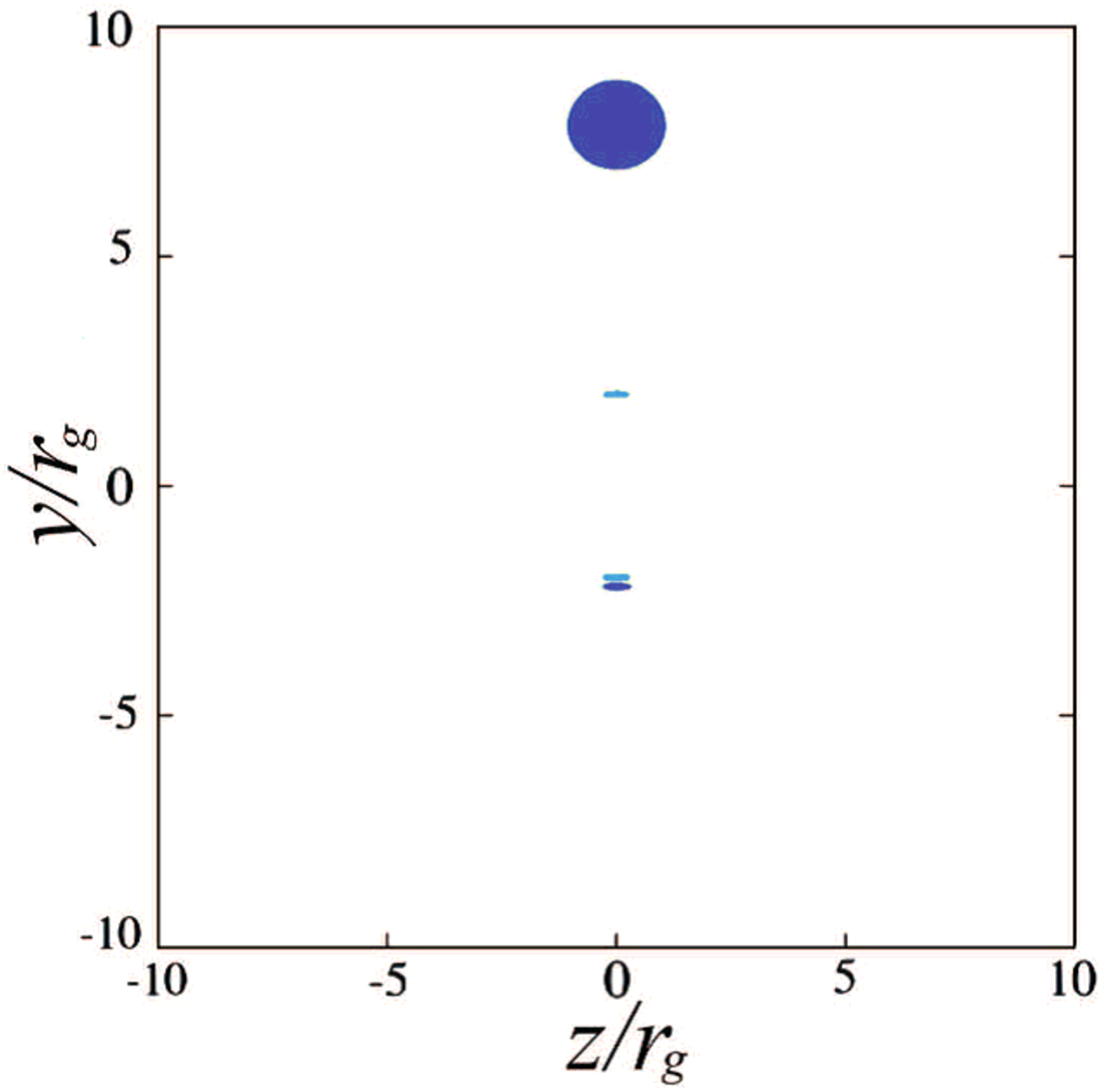}
\label{B130}}
\subfigure[$\varphi$=$150^\circ$]{
\includegraphics[scale=0.23]{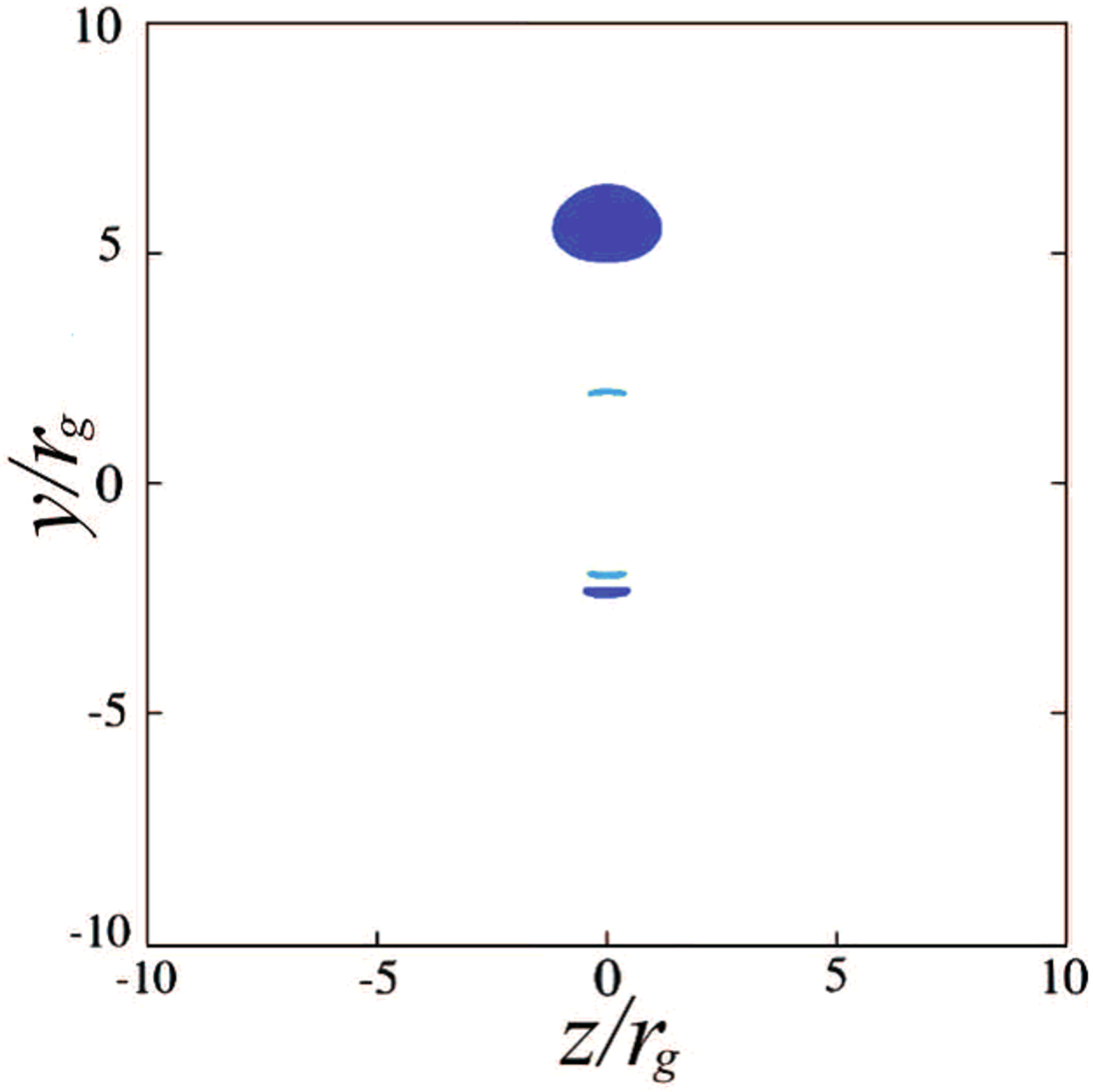}
\label{B150}}
  \subfigure[$\varphi$=$170^\circ$]{
\includegraphics[scale=0.23]{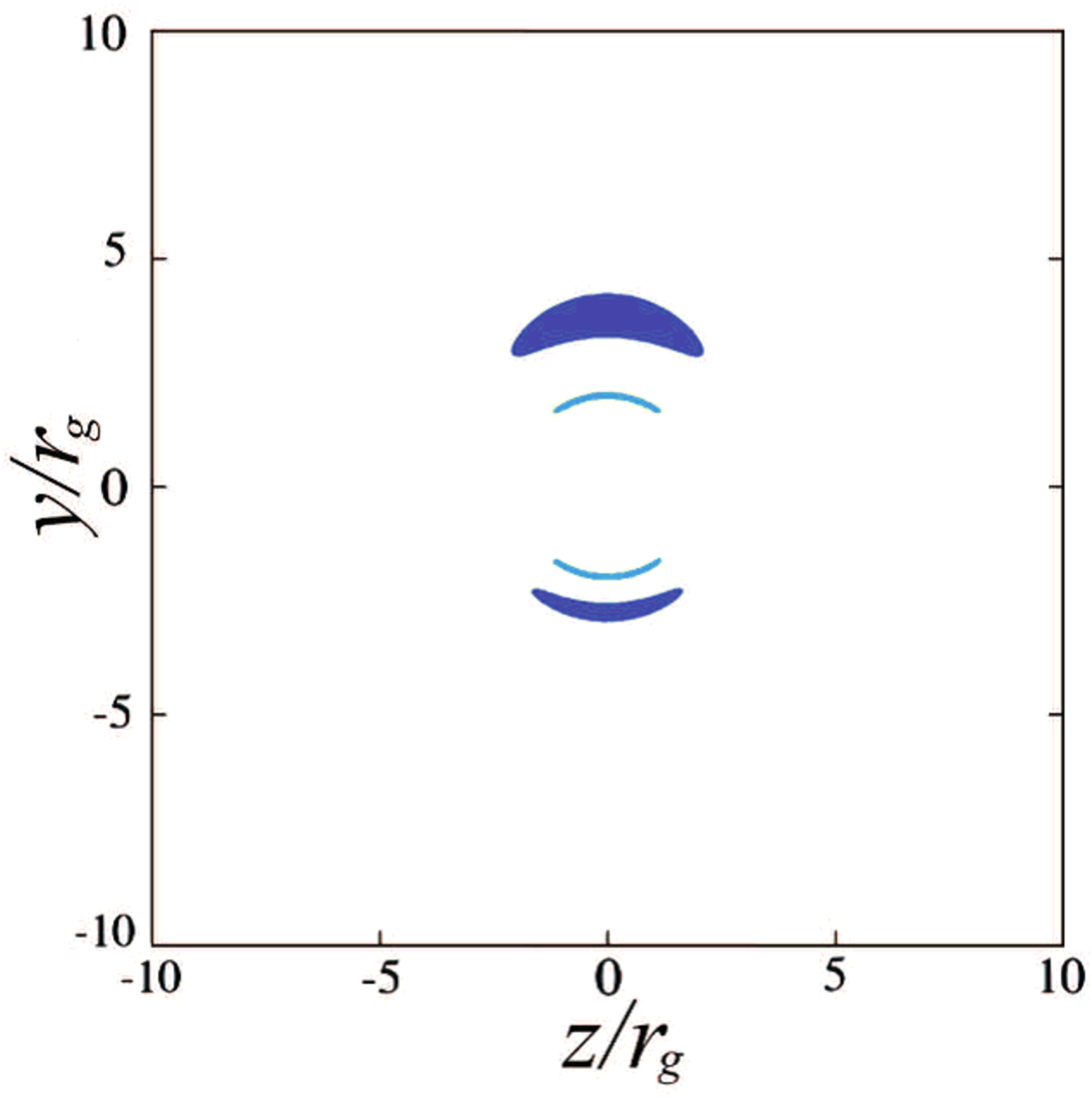}
\label{B170}}
\subfigure[$\varphi$=$180^\circ$]{
\includegraphics[scale=0.23]{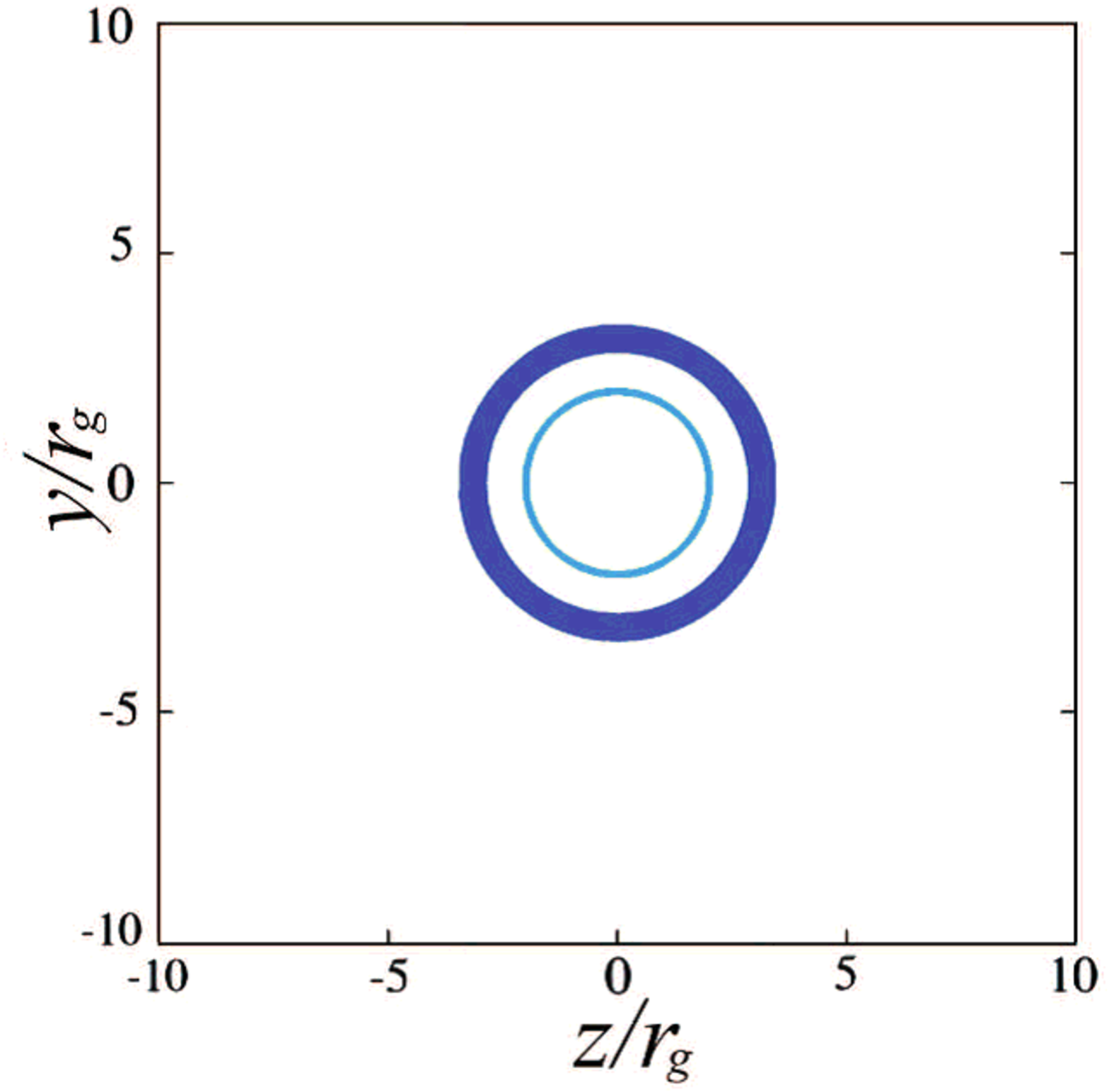}
\label{B180}}
\end{center}
\caption{Example of the gravitational lens images of the source object $\vec\alpha=(\alpha_y,~\alpha_z)$ around the braneworld black hole.
We set $D_o=300r_g,~D_s=10r_g$, and $r_s=r_g$.
We display four snapshots when $\varphi_s$=$130^\circ, 150^\circ, 170^\circ$, and $180^\circ$.
}
\label{br}
\end{figure}

\begin{figure}[H]
\begin{center}
\subfigure[$\varphi$=$130^\circ$]{
\includegraphics[scale=0.23]{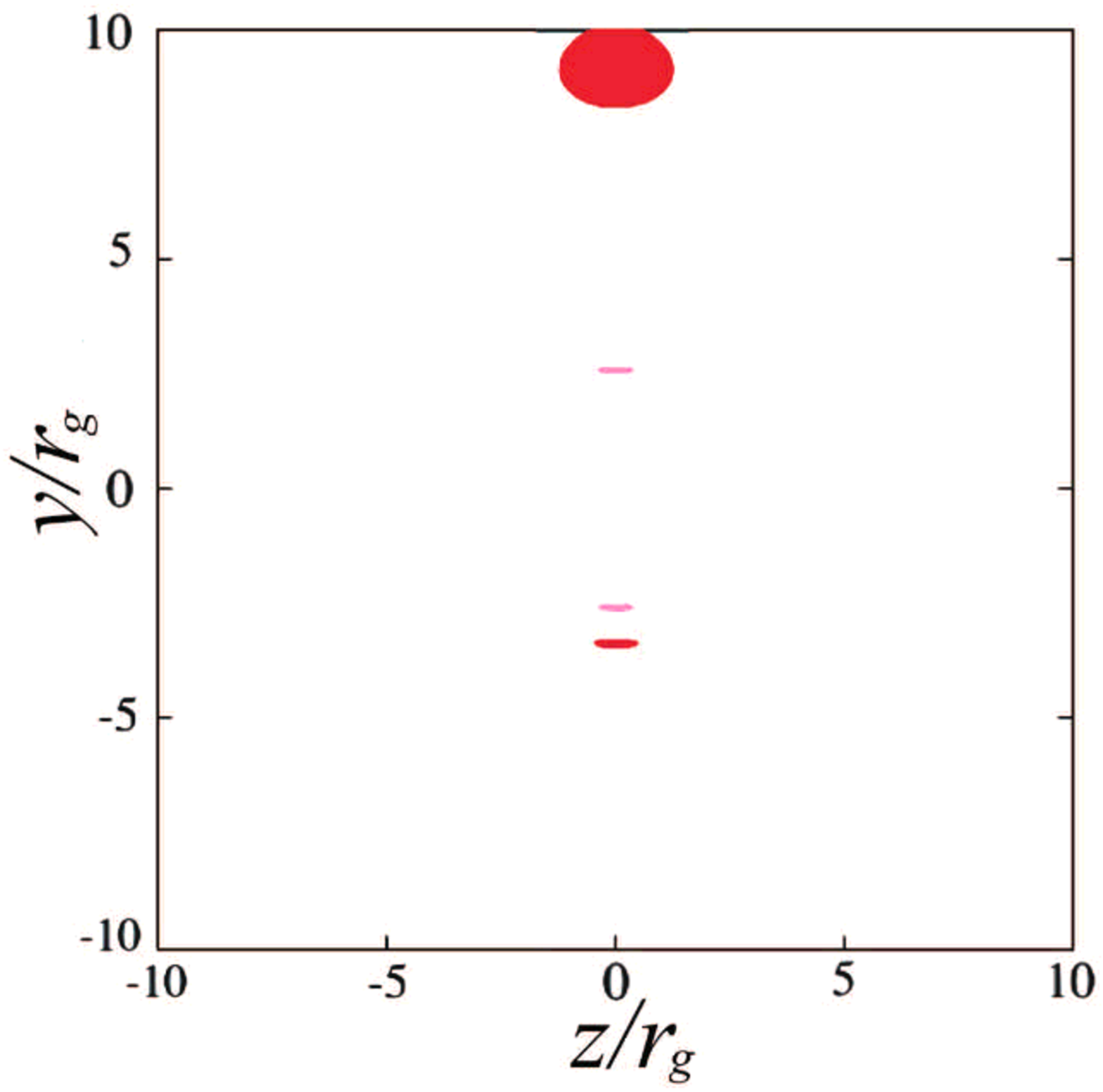}
\label{S130}}
\subfigure[$\varphi$=$150^\circ$]{
\includegraphics[scale=0.23]{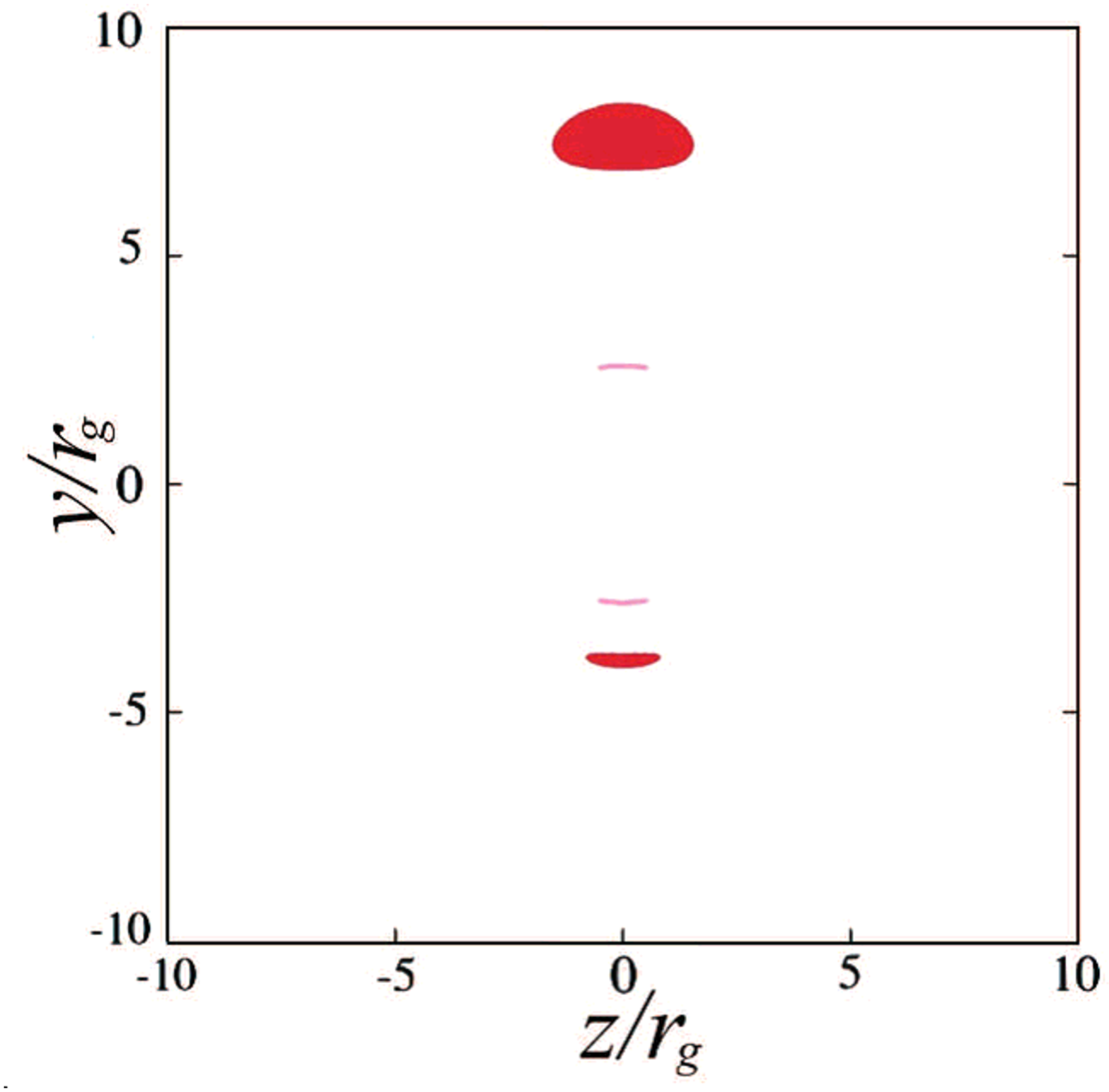}
\label{S150}}
\subfigure[$\varphi$=$170^\circ$]{
\includegraphics[scale=0.23]{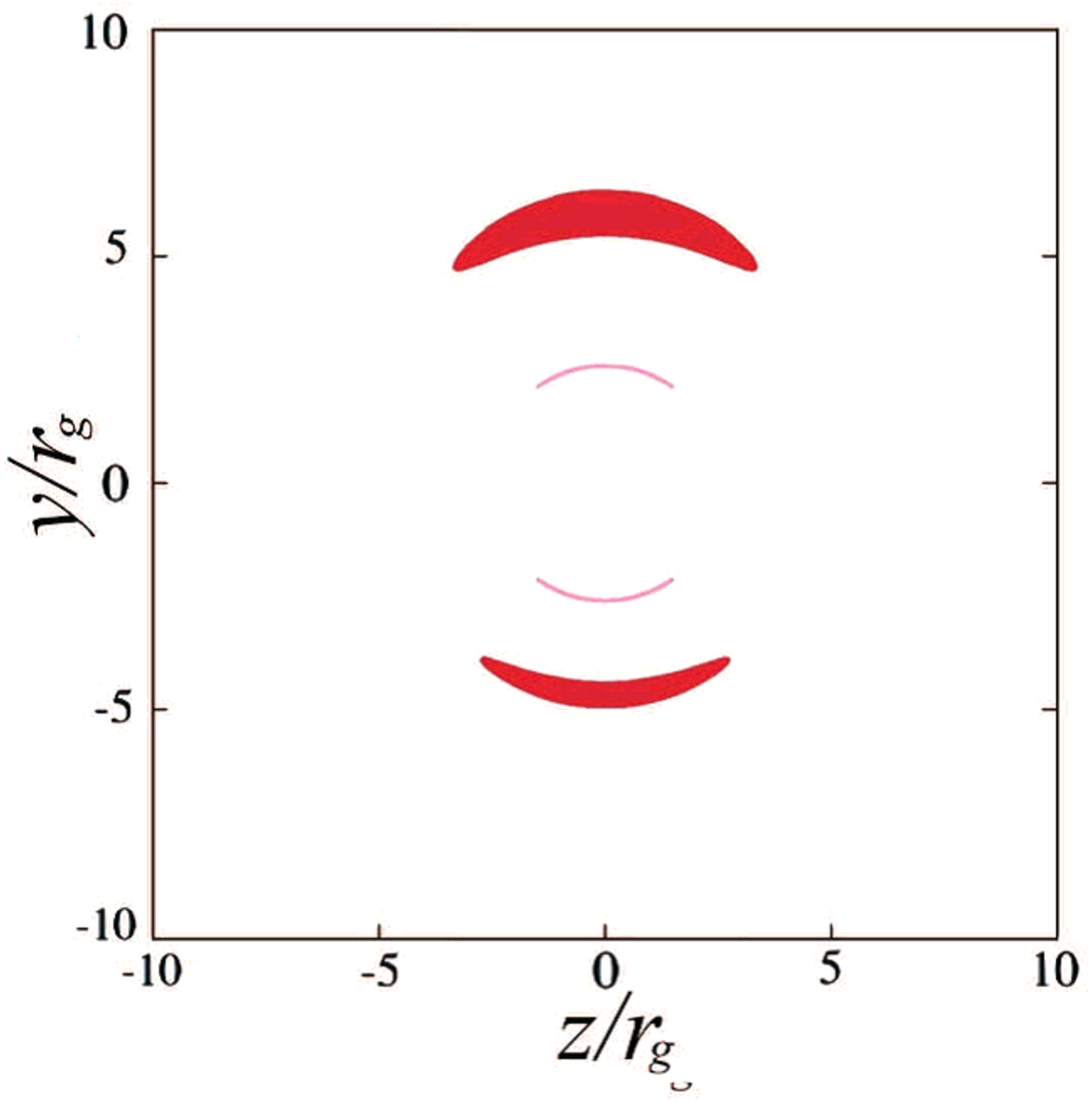}
\label{S170}}
\subfigure[$\varphi$=$180^\circ$]{
\includegraphics[scale=0.23]{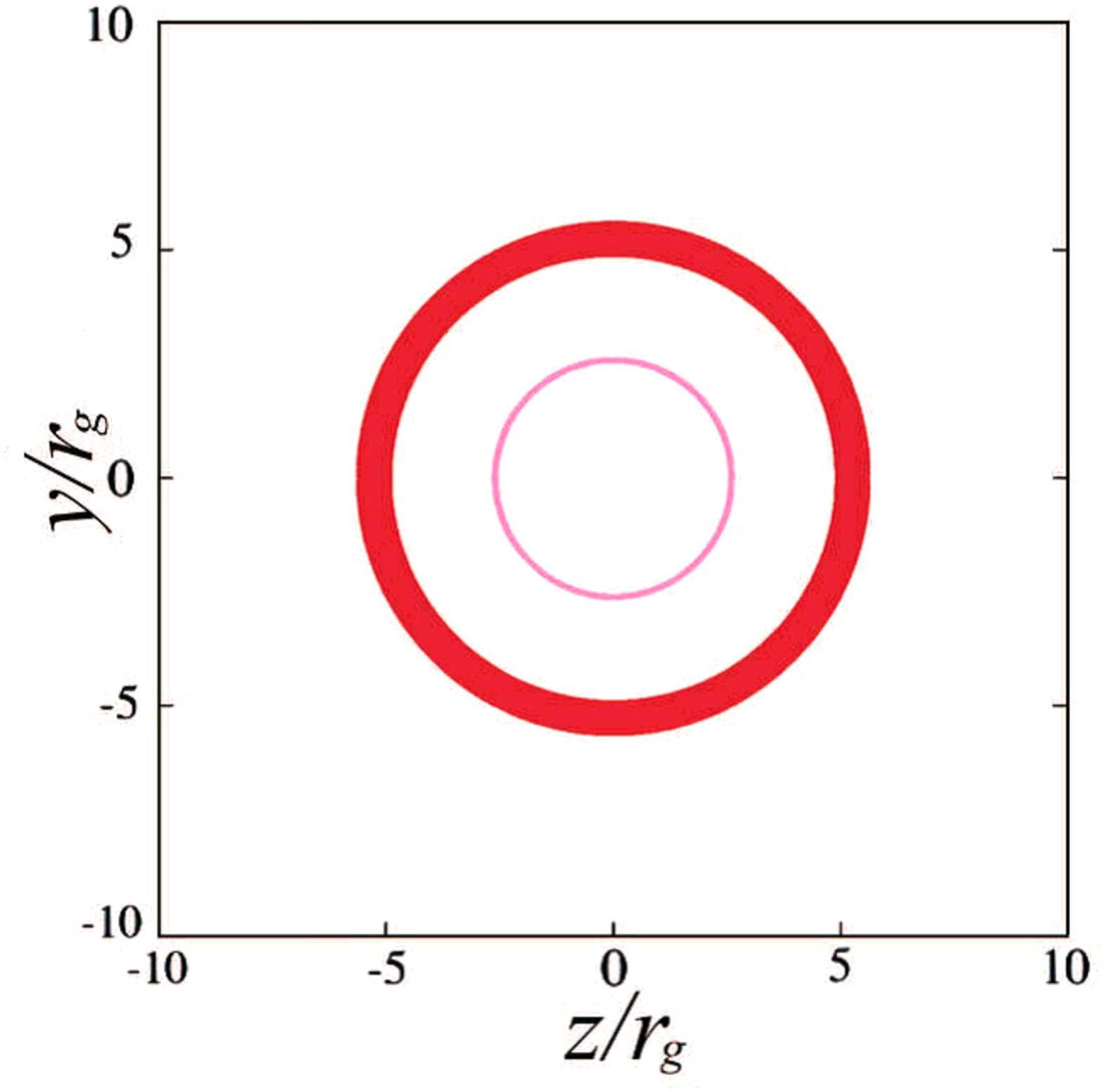}
\label{S180}}
\end{center}
\caption{The same as Fig.\ \ref{br}, but for the Schwarzschild black hole.}
\label{sr}
\end{figure}

\begin{figure}[H]
\begin{center}
\subfigure[$\varphi$=$130^\circ$]{
\includegraphics[scale=0.23]{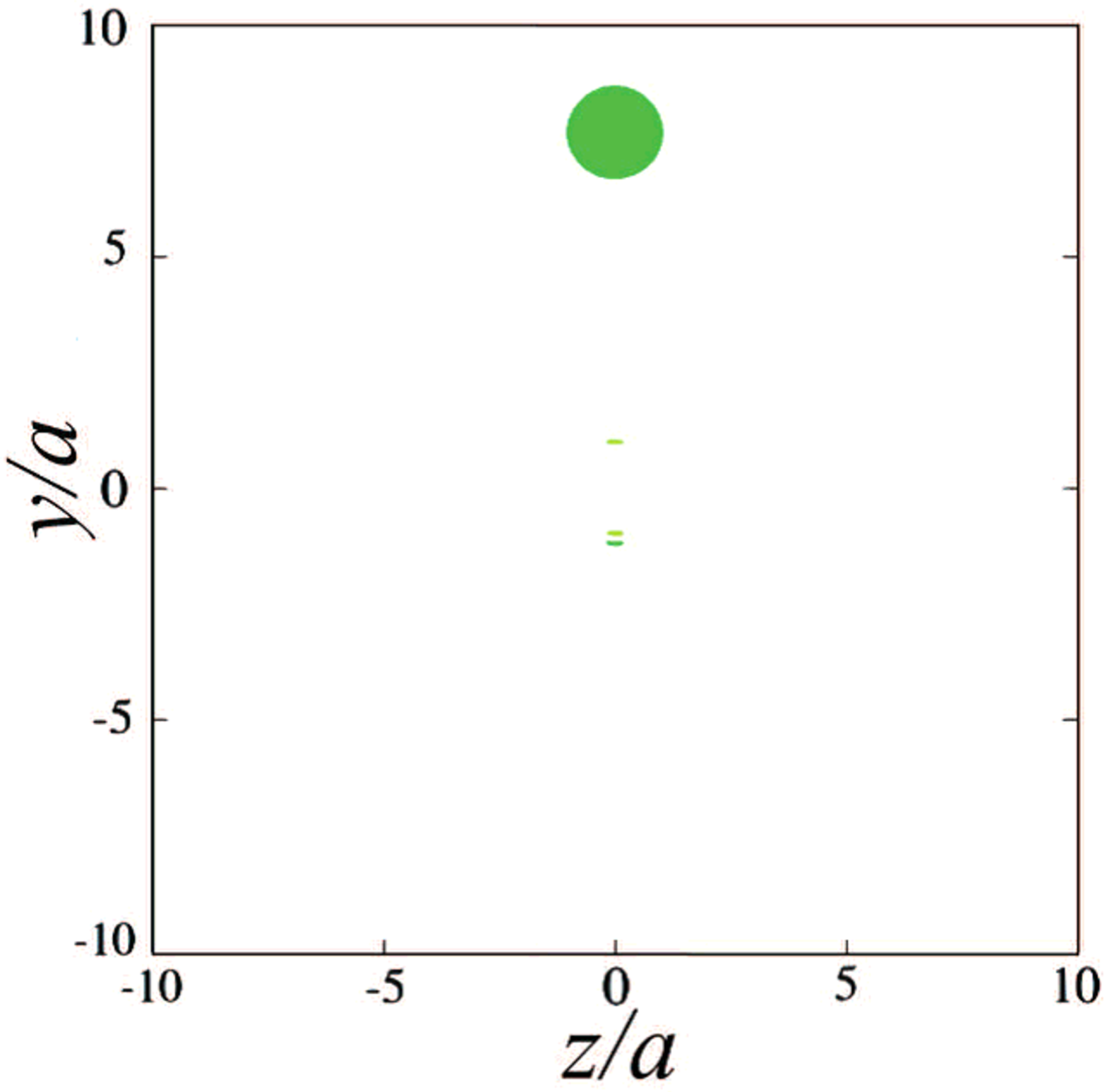}
\label{E130}}
\subfigure[$\varphi$=$150^\circ$]{
\includegraphics[scale=0.23]{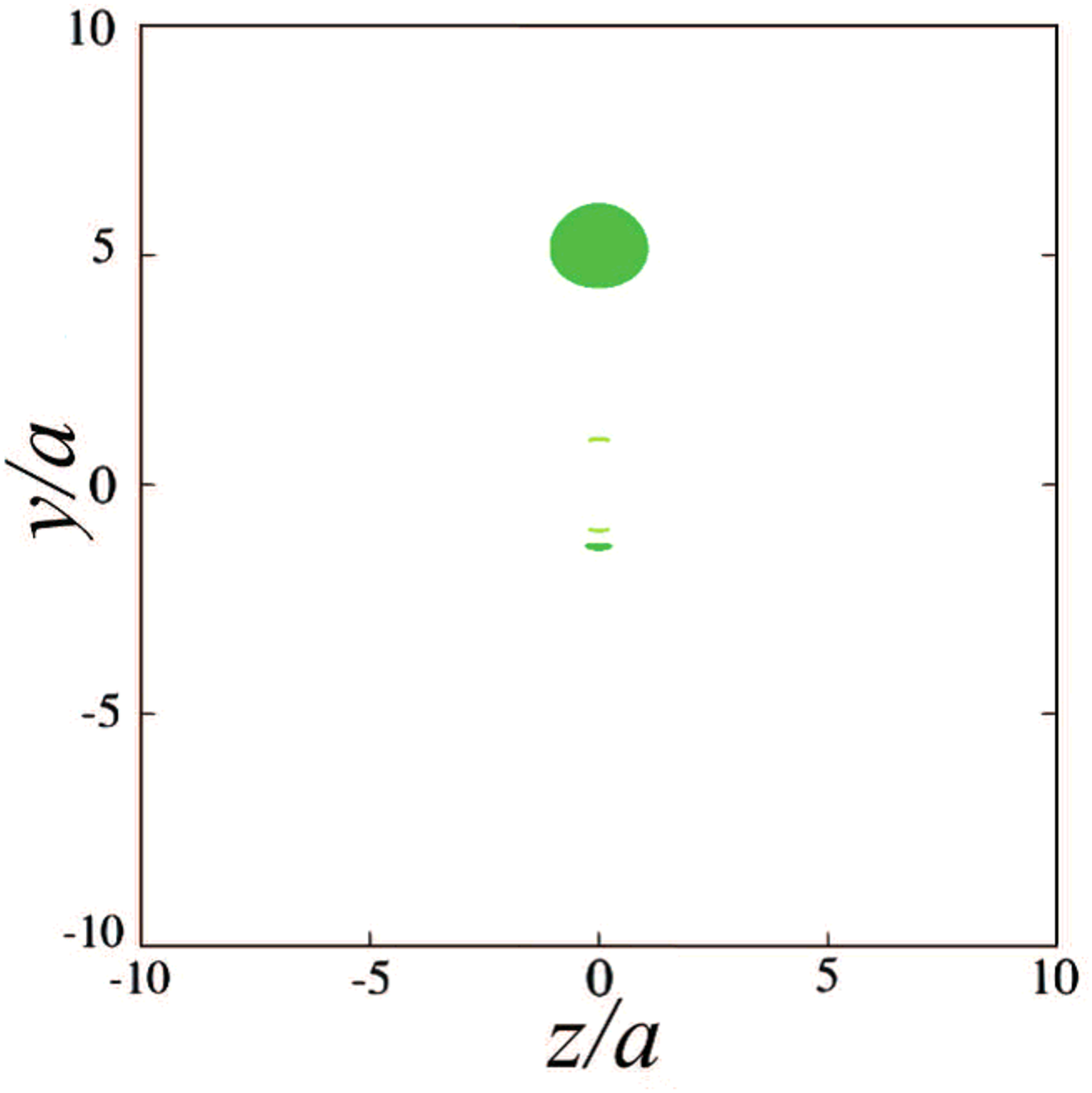}
\label{E150}}
\subfigure[$\varphi$=$170^\circ$]{
\includegraphics[scale=0.23]{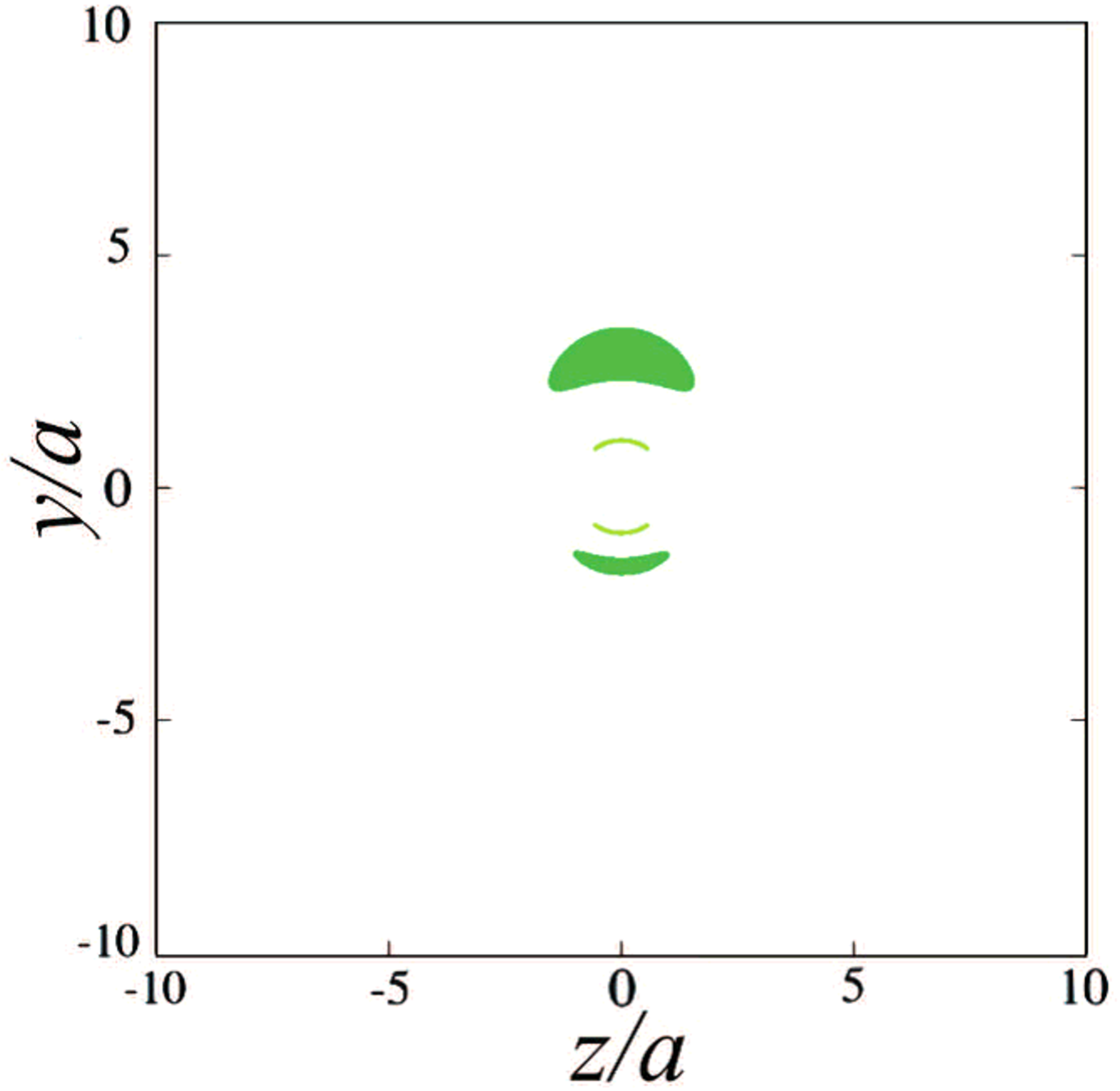}
\label{E170}}
\subfigure[$\varphi$=$180^\circ$]{
\includegraphics[scale=0.23]{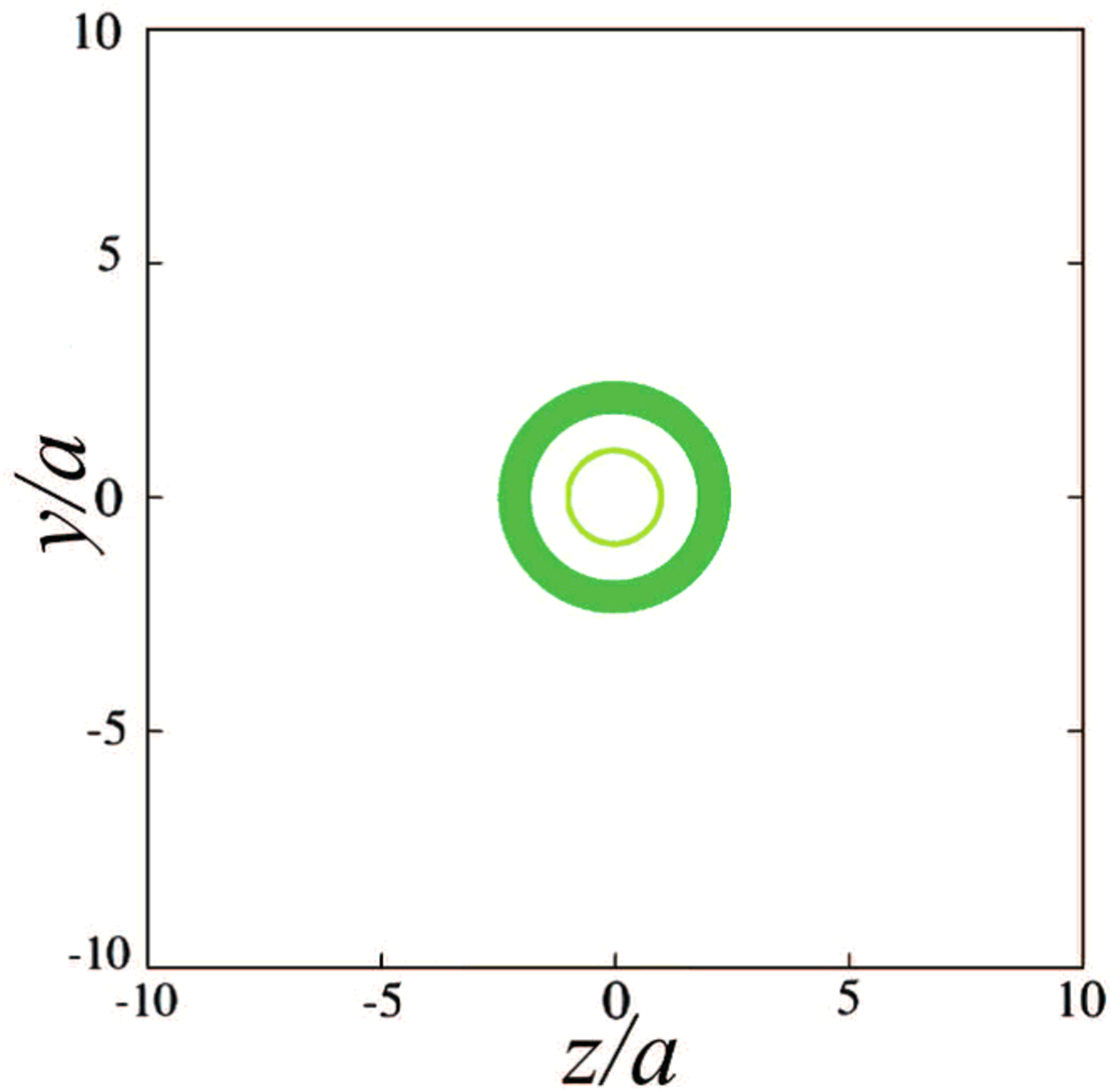}
\label{E180}}
\end{center}
\caption{The same as Fig.\ \ref{br}, but for the Ellis wormhole. We set $D_o=300a,~D_s=10a$, and $r_s=a$.}
\label{er}
\end{figure}

For reference, we also show gravitational lens images for the Schwarzschild black hole and for the Ellis wormhole in Figs.\ 6 and \ 7, respectively.
In Fig.\ 8, we superpose the three types of images when $\varphi$=$170^\circ$.
The qualitative features of the deformed images are common though the rate of deformation and amplification depends on lens objects.

\begin{figure}
\centering
\includegraphics[scale=0.26]{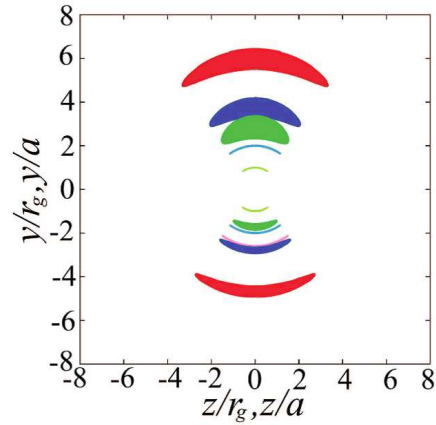}
\caption{Superposition of the gravitational lens images for the braneworld black holes (blue), the Schwarzschild black holes (red), and the Ellis wormholes (green) when $\varphi$=$170^\circ$.}
\label{image170}
\end{figure}

\begin{figure}
	\centering
	\includegraphics[scale=0.35]{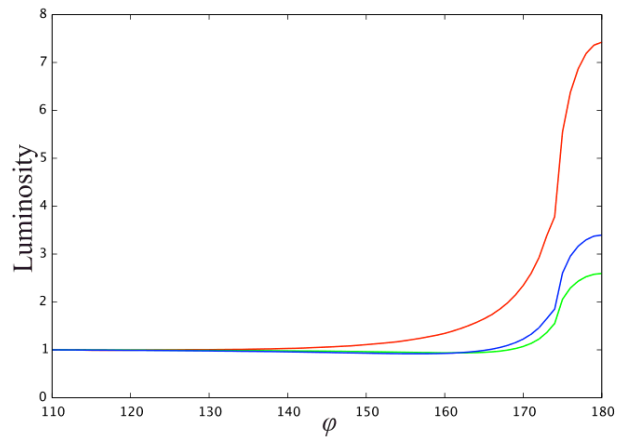}
	\caption{Light curves for the three models: the braneworld black holes (blue), the Schwarzschild  black holes (red), and the Ellis wormholes (green).
The ordinate denotes the magnification normalized by the value without lensing.
We assume that the gravitational redshift is negligible; then, the total magnification is proportional to the apparent area.
Amplification occurs for all cases; there is a peak at $\varphi=180^\circ$.
Gutters appear (i.e., extinction)  just before (and after) the amplification for the braneworld black hole and the Ellis wormhole.}
	\label{koudo3}
\end{figure}

To see how the intensity of the source object is changed by the lens objects, we plot light curves for the three models in Fig.\ 9.
We assume that gravitational redshift is negligible; then, the total intensity is proportional to the apparent area.
Amplification occurs for all cases; there is a peak at $\varphi=180^\circ$.
We also find that gutters appear (i.e., intensity reduction) just before (and after) the amplification for the braneworld black hole and the Ellis wormhole.
Therefore, if such reduction is observed in microlensing phenomena, the lens object would be either a braneworld black hole or a wormhole.

Because we normalize the scale by $r_g$ or $a$, the results in Figs. 8 and 9 are independent of $r_g$ or $a$. However, luminosity contract depends on the distance between the source and the lens, which is difficult to estimate in observations. Therefore, it is difficult to distinguish the braneworld black hole from the Ellis wormhole by comparing values of luminosity contract in microlensing phenomena solely. Thus, in the next section we consider another observational method.

\section{Stationary Dust flows and Shadows}

To distinguish between the braneworld black hole and the Ellis wormhole, we next consider stationary dust flows surrounding them and investigate their optical images, or shadows. 

\subsection{Stationary solutions of dust flows}

For the three spacetime models, we consider spherically symmetric and stationary dust flows.
Let $\rho(r)$ and $u^\mu(r)$ be the energy density and the $4$-velocity of dust, respectively;
then its energy-momentum tensor is given by
\beq
T^{\mu\nu}=\rho(r)u^{\mu} u^{\nu},
\eeq
where the $4$-velocity is expressed as
\beq
u^{\mu}=\frac{dx^{\mu}}{d\tau}=\left(u^t(r),u^r(r),0,0\right),
\eeq
and satisfies
\begin{equation}
-1=u_{\mu}u^{\mu}=-A\left(u^t\right)^2+B\left(u^r\right)^2.
\end{equation}
Here, we suppose ingoing flows, $u^r<0$, and therefore define $u(r)\equiv-u^r>0$.
The conservation law of energy momentum,
\begin{equation}\label{EMC}
T^{\mu \nu}_{~~;\nu}=0,
\end{equation}
 is concretely written as
\bea\label{eq-u}
&&{du\over dr}u -\frac{1}{2}{dA\over dr}=0,\\
&&\frac{d}{dr}\left(\rho r^2 u\right)=0.
\label{eq-rho}\eea

For the three models, we obtain the solutions of (\ref{eq-u}) and (\ref{eq-rho}) as follows:
\begin{itemize}
\item Schwarzschild black hole:
\bea
u&=&u_g\sqrt{\frac{r_g}{r}},\\
\rho&=&\rho_g\left(\frac{r_g}{r}\right)^{\frac{3}{2}}.
\eea
\item Braneworld black hole:
\bea
u&=&u_g\frac{r_g}{r},\\
\rho&=&\rho_g\frac{r_g}{r}.
\eea
\item Ellis wormhole \cite{OS}:
\bea
u&=&{\rm constant},\\
\rho&=&\rho_a\frac{a^2}{r^2 +a^2}.
\eea
\end{itemize}

Here, $\rho_g,~\rho_a,~u_g,$ and $u_a$ are integration constants.
Intensity contrast, which we will calculate below, does not depend on $\rho_g$ or $\rho_a$.
We set $u_g=-1$ because any particle falls in with light velocity at the event horion.
As for the constant velocity $u$ in the Ellis wormhole, we choose $u=10^{-7}$ as a typical value of interstellar matter velocity.

\subsection{Shadows}

To obtain the optical images of the lens object surrounded by dust, we compute the relationship between the apparent position of the optical source $\alpha$  and the intensity.
We suppose a gravity source and an observer and spread dust as shown in Fig. \ref{S&P&O}.
As we discussed in Sec.\ III.C, the impact parameter $\alpha$ is approximated by $L/E$ if the observer is sufficiently far from the object, and it indicates the apparent position of the optical source.

\begin{figure}[b]
	\centering
	\includegraphics[scale=0.34]{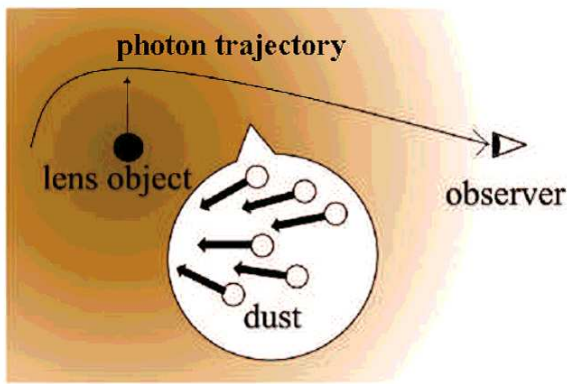}
	\caption{Setup of our analysis for obtaining optical images.
	We put an observer, a gravity source,  and dust surrounding it.
	The dust falls into the gravity source constantly. }
	\label{S&P&O}
\end{figure}

The radiation transfer equation is 
generally expressed as
\begin{equation}
	\frac{d J}{d \lambda}=\frac{\eta\left(\nu\right)}{\nu^2}-\nu \chi\left(\nu\right)J,~~~
	J\equiv{I(\nu)\over\nu^3},
	\label{eq: transfer equation}
\end{equation}
where $I(\nu)$ is the specific luminosity, $J $ is the invariant intensity, $\eta(\nu)$ is the emission coefficient, and $\chi(\nu)$ is the absorption coefficient  \cite{yusou}.
$\nu$ is the frequency measured by observers comoving with dust particles:
\begin{equation}
	\nu=-k_{\mu}u^{\mu}.
\end{equation}

Here, we make the following assumptions for simplicity:
\begin{itemize}
\item[(a)] The dust does not absorb radiation; i.e., $\chi(\nu)=0$.
\item[(b)] $\eta(\nu)$ is proportional to dust density, which is measured along null geodesics; i.e., $\eta(\nu)d\lambda\propto\rho u_\mu dx^\mu$.
\item[(c)] $\eta(\nu)$ is independent of $\nu$; i.e., the spectrum is flat.
\end{itemize}
The assumptions (b) and (c) indicate that $\eta(\nu)$ is expressed as
\begin{equation}
	\eta\left(\nu\right)=-H\left(\nu\right)\rho u_{\mu} k^{\mu},
\end{equation}
where $H$ is a constant. 
If the dust spectrum is not flat, $H$ should be a function of $\nu$.
Under these assumptions, we can rewrite (\ref{eq: transfer equation}) as an integral form,
\begin{equation}
	J=-\int\frac{H}{\nu^2}\rho u_{\mu}dx^{\mu}.
	\label{eq:integration of transfer equation}
\end{equation}

For the three spacetime models, we numerically calculate the luminosity distribution as follows:
\def\theenumi{(\roman{enumi})}
\begin{enumerate}
	\item Put the observer at $r=500 r_g$.
	\item For a given value of $\alpha=L/E$, we solve the null geodesic equations from the observer.  We can choose a value of the initial (observed) frequency $\nu_o$ arbitrarily because the ratio of $\nu_o$ to the emitted frequency $\nu_e$ does not depend on $\nu_o$.
	\item With the values of $\nu$ at each point, which are determined by the null geodesic equations, we integrate (\ref{eq:integration of transfer equation}) to obtain the luminosity $I$. We adopt the fourth-order Runge-Kutta method for all integrations.
	\item We continue the integrations until $r=500 r_g$ again, where gas density is sufficiently small.
	\item Iterate (ii) $\sim$ (iv) by changing the value of $\alpha$.
\end{enumerate}

We show our numerical results in Fig. \ref{kyoudobunpu}.
A luminosity peak caused by unstable circular orbits appears in all cases.
In the case of the Ellis wormhole, the inside of the peak $\alpha/{\alpha_{max}}<1$ is brighter than the outside $\alpha/{\alpha_{max}}>1$, contrary to the other cases. This is because of the existence of photons passing through the throat from the opposite side. We therefore speculate that this feature appears for other passable wormhole models. This difference would enable us to discriminate between braneworld black holes and Ellis wormholes by observations of shadows.

\begin{figure}[H]
	\centering
	\includegraphics[scale=0.35]{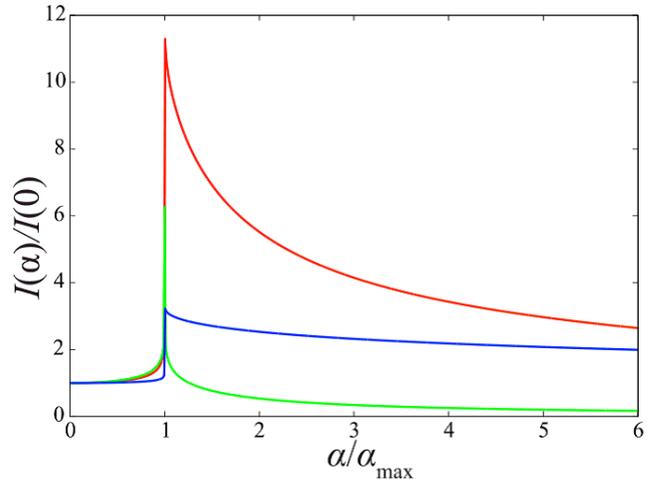}
	\caption{Numerical results of radiation luminosity for the Schwarzschild black hole (red),  the braneworld black hole (blue), and the Ellis wormhole (green).}
	\label{kyoudobunpu}
\end{figure}

\section{Conclusion}

We have investigated microlensing by massless braneworld black holes and their shadows and discuss how to distinguish them from standard Schwarzschild black holes and Ellis wormholes.
First, we studied defection angles of light rays that pass around those objects.
Previous work showed that both deflection
angles of the braneworld black hole and the Ellis wormhole are proportional to $\alpha^{-2}$,
while that of the Schwarzschild black hole to $\alpha^{-1}$. 
We therefore speculated that the braneworld black hole and the Ellis wormhole may exhibit similar features in microlensing phenomena.

To elucidate observational consequences of those microlensing phenomena, 
we calculated images of an optical source object behind a lens object for the three models and their light curves.
We found that for the braneworld black hole as well as for the Ellis wormhole, luminosity reduction appears just before and after amplification.
This means that, observations of such reduction would indicate the lens object is either a braneworld black hole or a wormhole, though it is difficult to distinguish one from the other by microlensing solely.

Thus, we next analyzed the optical images of the braneworld black hole surrounded by optically thin dust and compared them with those of the Ellis wormhole.
Because the spacetime around the braneworld black hole possesses unstable circular orbits of photons, a bright ring appears in the image, just as in Schwarzschild spacetime or in the wormhole spacetime.
This indicates that the appearance of a bright ring does not solely confirm a braneworld black hole, a Schwarzschild, nor an Ellis wormhole.
However, we found that only for the wormhole the intensity inside the ring is larger than the outsider intensity.
Our results mean that observations of shadows would distinguish black holes from Ellis wormholes.

We therefore conclude that, with future high-resolution very long baseline interferometry observations of microlensing and shadows together, we could identify the braneworld black holes if they exist.

\section*{Acknowledgment}
This work was supported by JSPS KAKENHI Grant No. JP17J00547.

\end{document}